\documentclass[twocolumn,showpacs,preprintnumbers,amsmath,amssymb,superscriptaddress, prl,floatfix]{revtex4-1}
\usepackage{graphicx}% Include figure files
\usepackage{dcolumn}% Align table columns on decimal point
\usepackage{tabularx}
\usepackage{bm}% bold math
\usepackage{epstopdf}
\usepackage{amsmath}
\usepackage{color}

\begin{document}

\preprint{Preprint}

\title{Heterodyne Spectroscopy of Polariton Spinor Interactions}% Force line breaks with \\

\author{N. Takemura}
\email[E-mail: ]{naotomo.takemura@epfl.ch}
\affiliation{Laboratory of Quantum Optoelectronics, \'Ecole Polytechnique F\'ed\'erale de Lausanne, CH-1015, Lausanne, Switzerland}
\author{S. Trebaol}
\affiliation{Laboratory of Quantum Optoelectronics, \'Ecole Polytechnique F\'ed\'erale de Lausanne, CH-1015, Lausanne, Switzerland}
\author{M. Wouters}
\affiliation{Theory of Quantum and Complex Systems, Universiteit Antwerpen, B-2610 Antwerpen, Belgium}
\author{M. T. Portella-Oberli}
\affiliation{Laboratory of Quantum Optoelectronics, \'Ecole Polytechnique F\'ed\'erale de Lausanne, CH-1015, Lausanne, Switzerland}
\author{B. Deveaud}
\affiliation{Laboratory of Quantum Optoelectronics, \'Ecole Polytechnique F\'ed\'erale de Lausanne, CH-1015, Lausanne, Switzerland}

\date{\today}% It is always \today, today,

\begin{abstract}
We report on spinor polariton interactions in GaAs based microcavities. This investigation is carried out by means of heterodyne polarized pump-probe spectroscopy. We show the dependence of the energy renormalization of the lower and upper polariton resonances with cavity detuning for different polariton densities. We use the exciton-photon based Gross-Pitaevskii equation to model the experiment for both lower and upper polariton modes. The theoretical results reproduce qualitatively the experimental observations revealing the magnitude and the sign of the parallel and anti-parallel spin interaction strength. We evidence the strong influence of the biexciton resonance on the anti-parallel spin polariton energy shift and provide the exciton-biexciton coupling constant. We derive our results in the lower polariton basis using Gross-Pitaevskii equation from which, we express analytically the spinor polariton interactions and identify the clear role of the biexciton resonance.
\end{abstract}

%68.35.-p 	Solid surfaces and solid?solid interfaces: structure and en3333ergetics
%68.37.Ma 	Scanning transmission electron microscopy (STEM) 
%79.20.Uv 	Electron energy loss spectroscopy

\maketitle
\section{I. INTRODUCTION}
Microcavity exciton-polaritons are quasiparticles re-sulting from the strong coupling between excitons and photons \cite{Weisbuch1992}. Polaritons exhibit mutual interactions coming from their excitonic content and light effective mass inherited from the photon. Collected photons emitted from the cavity allow reading out the polaritons properties. As a matter of fact, a polariton fluid  is an ideal tool for investigating quantum phenomena in solid-state systems. Polariton interactions in semiconductor microcavities play a crucial role in a wide variety of topics ranging from nonlinear optical effects \cite{Savvidis2000,Messin2001,Saba2001,Kundermann2003,Langbein2004}, polariton superfluidity \cite{Amo2009,Utsunomiya2008,Kohnle2011, Kohnle2012} to Bose Einstein condensation \cite{Kasprzak2006}.\\
All these topics enlighten that polaritons provide a concrete realization of a many-body interacting system. As polaritons carry a spin, spinor interactions characterize fundamental physical processes in polariton quantum systems. This results in anisotropic nonlinearities at the origin of many effects such as stimulated spin dynamics of polaritons \cite{Lagoudakis2002,Kavokin2003}, the transport of spin polarized polaritons \cite{Langbein2007}, the optical spin Hall effect \cite{Kavokin2005,Leyder2007,Amo2009a}, the generation of polarization vortices \cite{Liew2007,Liew2008} and half quantum vortices \cite{Rubo2007}, spontaneous polarization buildup in Bose-Einstein condensation \cite{Baumberg2008}, in bistability \cite{Baas2004} multistability \cite{Gippius2007,Paraiso2010,Wouters2013} and polariton switching \cite{Amo2010,Cerna2013}. \\
Despite their importance, the spinor polariton interactions have been determined only indirectly. Several experiments indicate that the interaction of polaritons with anti-parallel spins is attractive \cite{Kuwata-Gonokami1997,Kavokin2005a,Vladimirova2010}. In others, with the presence of a reservoir it appears to be repulsive \cite{Paraiso2010, Wouters2013, Amo2010,Ferrier2011}. Furthermore, theoretical works predict that spinor polariton interaction strengths depend on the cavity detuning \cite{Kwong2001,Wouters2007,Vladimirova2010}. It is very important to note that each investigation uses its own experimental condition: resonant or non-resonant excitation, different cavity detuning and time scale, which influences in the measurement and then in the determination of the interaction constants. Furthermore, previous studies concentrate on the lower polariton behavior omitting the upper polariton branch.\\
In this work, we report on the study of lower and upper polariton energy renormalization in function of the polariton density and the exciton-photon cavity detuning in semiconductor microcavity. To carry out this investigation, we concentrate on the lower and upper polariton resonances through spectrally resolved pump-probe spectroscopy. We employ pump-probe pulse co- and counter-circular polarization configurations in order to investigate polariton-polariton interactions with parallel and anti-parallel spins, respectively. The results reveal either the repulsive or attractive character of the spinor polariton interactions through the measured energy shift of the probed lower and upper polariton resonances. Furthermore, the amplitude and sign of the lower and upper polariton energy shift as a function of the cavity detuning are determined for different polariton densities. Our theoretical model is developed in the framework of the spinor Gross-Pitaevskii equation in the exciton-photon basis. Comparison between numerical simulations and experimental measurements allows to extract the microscopic spinor interaction constants and also the strengh of the exciton-biexciton coupling constant. The biexciton constant evidences the strong influence of the biexciton resonance on the anti-parallel spin energy renormalization. Through basis transformation from exciton-photon to lower polariton, we derive analytical expressions for the spinor polariton interaction constants, usually referred to as $\alpha_1$ and $\alpha_2$ \cite{Ciuti2000}.\\
This paper is organized as follows: in Section II, we describe the sample and the pump-probe experiment. Section III reports on the experimental results, while Section IV is dedicated to the theoretical simulations using exciton-photon based Gross-Pitaevskii equations. In Section V, we identify analytically the spinor polariton interactions using polariton basis Gross-Pitaevskii equation. We then give a general conclusion in Section VI.
\section{II. EXPERIMENTAL METHOD}
This study is performed with a high quality III-V GaAs-based microcavity \cite{Stanley1994}. A single 8 nm In$_{0.04}$Ga$_{0.96}$As quantum well is introduced between a pair of GaAs/AlAs distributed Bragg-reflectors. The exciton-cavity detuning energy $\delta$ can be adjusted by changing the position of the laser spot on the wedged sample. The Rabi splitting is $\Omega$=3.26 meV at $\delta =0$. We use a pump-probe setup with both pump and probe 125 femtosecond pulses from a Ti:Sapphire laser. The center energy of the laser spectrum (the linewidth is 14.6 meV) is set between lower and upper polaritons, therefore the laser intensity is fixed for all cavity detunings. The experiments are performed under resonant excitation and at 4K.\\
We employ a heterodyne pump-probe technique \cite{Hall1992}. The laser beam is split into three: The pump and probe are frequency shifted with acousto-optic modulators (AOM) by 75 MHz and 79MHz, respectively, and focused on the sample. In transmission, the probe signal is directed into an AOM together with the reference beam. The AOM, driven at 79 MHz, produces two $\pi$-shifted detection channels in which the reference and probe signal overlap spectrally. The mixed beams are dispersed in a spectrometer and we subtract the two $\pi$-shifted interferograms to recover the pump-probe signal. This heterodyne pump-probe technique allows us to study the polariton interactions close to degenerate beams configuration at k = 0 and dramatically increases the signal-to-noise ratio. This enables a precise measurement of small energy shifts in the probe spectrum. The experimental setup is described in detail in our previous paper \cite{Kohnle2012}.\\
The principle of the experiment is the following: We generate a spin up lower and upper polariton population with a $\sigma^+$ circularly polarized pump pulse. By varying the power of the pump pulse we control the spin-up polariton population. A small amount of spin-up or -down lower and upper polaritons is then injected with a weak $\sigma^+$ or $\sigma^-$ circularly polarized probe pulse. We spectrally probe the energy of the lower and the upper polariton resonance by measuring the transmission spectrum of the probe pulse.\\
The strength of the spinor polariton interaction is determined from a renormalized energy shift of the polariton resonance due to the presence of the pump polaritons. If the spinor polariton interaction is repulsive, it shows a blueshift, otherwise a redshift appears for attractive interaction. The experiments are performed at different cavity detunings to highlight the role of the excitonic content on the polariton-polariton interactions.
\section{III. EXPERIMENTAL RESULTS}
\begin{figure}[t]
\begin{center}
\includegraphics[width=0.49\textwidth]{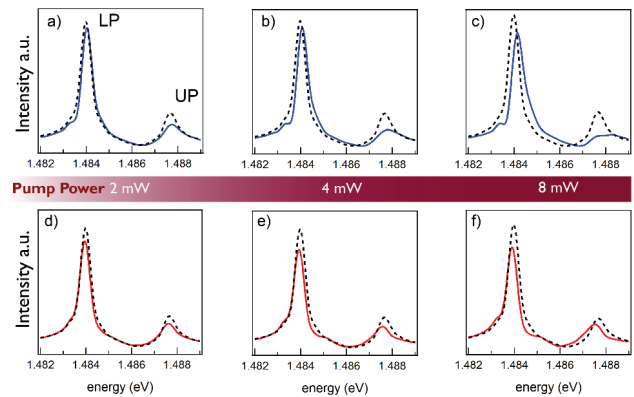}
\caption{\label{fig1} 
(Color online) Co- (blue) and counter- (red) circular polarization pump-probe spectra at -1.5 meV cavity detuning for different pump intensities: $3.0\times 10^{13}$ (a,d), $5.9\times 10^{13}$ (b,e) and $1.2\times 10^{14}$ (c,f) photons pulse$^{-1}$ cm$^{-2}$. Transmitted probe spectra without pump (black dashed lines) and with pump (solid lines).}
\end{center}
\end{figure}

In Figure \ref{fig1}, we display the co- and counter-circular pump-probe spectra measured at -1.5 meV cavity detuning for pump photon density of $3.0\times 10^{13}$, $5.9\times 10^{13}$ and $1.2\times 10^{14}$ photons pulse$^{-1}$cm$^{-2}$ together with the reference spectrum.\\
The pump-probe spectra clearly show a blueshift of both the lower and upper polariton resonances in co-circular configuration, while for counter-circular measurements they show a redshift. These results also show that, as the polariton density is increased, the polariton resonance energy shift increases. In order to extract the energy shifts of the polariton resonance quantitatively, each spectrum is fitted with a Lorentzian function. At a fixed cavity detuning and for each pump power, the energy shift is determined through the difference between the reference and the pump-probe spectra.\\
We analyze mutually the lower and upper polariton energy shift for different polariton density and cavity detuning. We investigate separately the parallel and anti-parallel spin polariton interactions. 
\subsection{A. Parallel spin polariton interaction}
\begin{figure}[t]
\begin{center}
\includegraphics[width=0.4\textwidth]{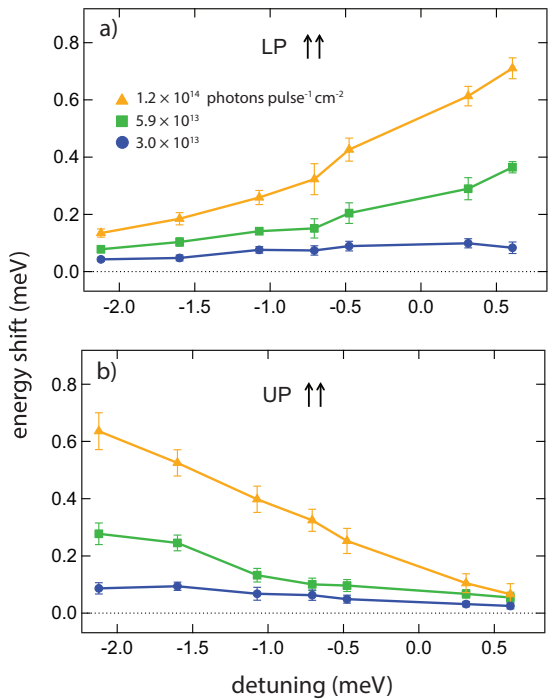}
\caption{(Color online) Energy shifts of lower (a) and upper (b) polariton resonances for co-circularly polarization configuration as a function of the cavity detuning. Three different symbols represent different pump intensities.}\label{co}
\end{center}
\end{figure}

In Figure \ref{co} (a), we show together the dependence of the lower and upper polariton energy shifts with cavity detuning, for different polariton densities. The results show that the blueshifted for both resonance increases with polariton density and varies with cavity detuning. This evidences the repulsive interaction between polariton with parallel spin for all cavity detuning. Notice the “mirror” behavior of the upper-lower polariton energy shift with cavity detuning. This reflects the role of the excitonic content of polariton states on polariton-polariton interactions. Indeed, for negative detuning the upper polariton being more excitonic-like than lower polariton, presents larger blueshift. The reverse is valid for positive detuning.\\
%Second, we stress that the mirror symmetry is not at zero detuning. Indeed, for a given pump photon density, upper polariton state experiences an energy shift comparable to lower polariton state for a larger absolute detuning value. We attribute this behavior to the onset of relaxation that depopulate upper polariton state into the excitonic reservoir through polariton-phonon interaction. Therefore, the depopulation of the upper polariton mode induces a reduction of the energy shift which eventually prevents the weak coupling transition to occur \cite{footnote}. This phonon scattering process becomes efficient only at negative detuning when upper polaritons are more excitonic and for small energy separation between the upper polariton and the exciton reservoir \cite{Piermarocchi1996,Stanley1997,Grosso2014}. It is worth mentioning that the  reservoir density is small enough to prevent energy shift renormalization on polariton states.
% Excitations from the upper state are spread in the momentum space trough the scattering process. Therefore the density of states is relatively small do not 
We would like to mention that the mirror symmetry axis is not at zero detuning but appears around $\delta=-0.8$ meV. We might attribute this behavior to an onset of upper polariton scattering into an exciton reservoir. Actually, to compensate this depopulation, the upper polariton resonance needs to be more excitonic to experience energy shift comparable to the lower polariton.

\subsection{B. Anti-parallel spin polariton interaction}
\begin{figure}[t]
\begin{center}
\includegraphics[width=0.4\textwidth]{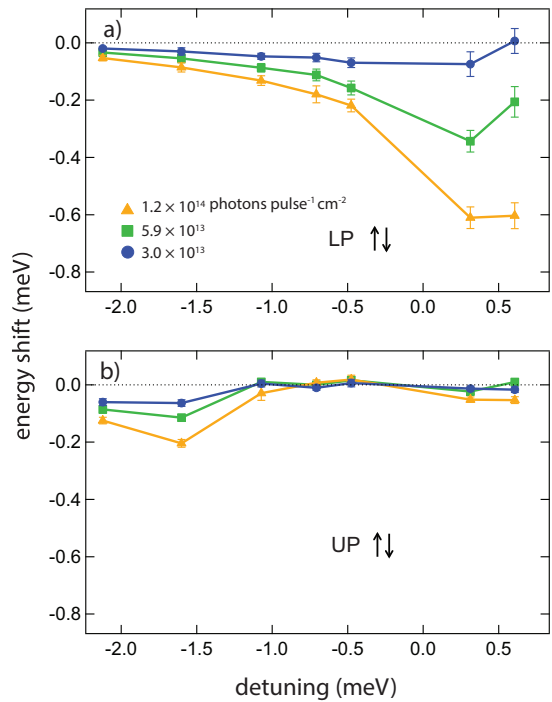}
\caption{(Color online) Energy shifts of lower (a) and upper (b) polariton resonances for counter-circularly polarization configuration as a function of the cavity detuning. Three different symbols represent different pump intensities.}\label{cross}
\end{center}
\end{figure}

For the same set of polariton densities as above, we display the dependence of the upper and the lower polariton energy shifts with cavity detuning in Figure \ref{cross}(a) and (b), respectively. The lower and upper polariton resonances are red shifted and this shift increases with polariton density. This result signs the attractive interaction between polariton with anti-parallel spin. The influence of the excitonic content of the polaritons for the interactions is again evidenced. Eventually, for lower polaritons, the energy shift increases from negative to positive cavity detunings. The reverse happens for the energy shift of the upper polaritons.\\
However, the dependence of the upper polariton energy shift with cavity detuning is not a “mirror” image of the lower polariton as for the results presented for parallel spins. Here, the lower polaritons undergo a much stronger energy shift than the upper polaritons. Actually, the lower polariton energy renormalization is not only due to polariton-polariton interaction but also due to polariton-biexciton coupling. It is important to note that for the lower polaritons from negative to positive cavity detuning, the energy of two-lower polaritons approaches the energy of the biexciton resonance. This effect highlights the influence of the biexciton on the lower polariton energy renormalization for anti-parallel spin polariton interactions.  This result is in agreement with the polaritonic Feshbach resonance behavior recently observed \cite{Takemura2014} corroborating that biexcitons play a crucial role in the anti-parallel spin polariton interactions.
 
\section{IV. THEORETICAL MODEL}
Since our experimental approach consists of the excitation of both lower and upper polariton states in the nonlinear regime, cross interactions between polariton modes might appear leading to an hazardous description of our results in the polariton basis. Therefore, we use the exciton-photon based Gross-Pitaevskii equation to reproduce the experimental observations. Comparing the simulation to the experimental data will allow us to extract the spinor interaction constants in the exciton-photon basis Hamiltonian. In section V, we will show how to transpose our analysis in the lower polariton basis yielding a comparison to previously reported results.\\
The exciton-photon basis Hamiltonian is the following:
\begin{equation}
\hat{H}=\epsilon_c\hat{c}^{\dagger}_{\uparrow}\hat{c}_{\uparrow}+\epsilon_x\hat{x}^{\dagger}_{\uparrow}\hat{x}_{\uparrow}+\epsilon_b\hat{B}^{\dagger}\hat{B}+\Omega(\hat{c}^{\dagger}_{\uparrow}\hat{x}_{\uparrow}+\hat{x}^{\dagger}_{\uparrow}\hat{c}_{\uparrow})+\hat{H}_{\rm int}.
\label{eq_H}
\end{equation}
The interacting part $\hat{H}_{\rm int}$ is given by
\begin{equation}\label{eq_H_cxb}
\begin{split}
\hat{H}_{\rm int}&=
\hat{H}_{\uparrow \uparrow} + \hat{H}_{\uparrow \downarrow} +\hat{H}_{bx}\\
&=
g_{\scriptscriptstyle ++}\hat{x}^{\dagger}_{\uparrow}\hat{x}^{\dagger}_{\uparrow}\hat{x}_{\uparrow}\hat{x}_{\uparrow} + g_{\scriptscriptstyle +-}\hat{x}^{\dagger}_{\uparrow}\hat{x}^{\dagger}_{\downarrow}\hat{x}_{\downarrow}\hat{x}_{\uparrow}\\
&+ g_{bx}(\hat{B}\hat{x}_{\uparrow}^{\dagger}\hat{x}_{\downarrow}^{\dagger}+\hat{x}_{\uparrow}\hat{x}_{\downarrow}\hat{B}^{\dagger})
\end{split}
\end{equation}
The $c$, $x$ and $B$ are the photon, exciton and biexciton annihilation operators, $\epsilon_x$, $\epsilon_c$ and $\epsilon_B$ are respectively the exciton, photon and biexciton energy. The arrow $\uparrow$ and $\downarrow$ defines the spin polarization as up and down. The $g_{\scriptscriptstyle ++}$ and $g_{\scriptscriptstyle +-}$ are respectively the interaction strengths for parallel and anti-parallel exciton interaction. The coupling between excitons and biexciton is given by $g_{ bx}$.\\
It might be worth commenting the microscopic origin of these phenomenological interaction constants. To be precise, excitons are composite bosons composed of electron-hole pairs, thus exciton-exciton interactions are governed by Coulomb interaction between electrons and holes. The simplest approach to calculate exciton-exciton interaction with parallel spins $g_{\scriptscriptstyle ++}$ is to calculate scattering matrix of excitons based on the Born approximation \cite{Ciuti1998}. According to this calculation, in zero momentum scattering, which is the case of our experiment, the main contribution to the exciton-exciton interaction is electron-hole exchange interaction. However, this calculation cannot explain the existence of the exciton interaction with anti-parallel spins, because the exchange interaction disappears between excitons between anti-parallel spins \cite{Ciuti1998}. Therefore, the inclusion of biexciton state ($g_{bx}$ in our model) and the calculation beyond the Born approximation are important to explain the exciton interaction with anti-parallel spins \cite{Axt1998,Kwong2001a,Takayama2002,Oestreich1995a}. For example, such a calculation has been done by \cite{Kwong2001a,Takayama2002} through the summation of higher orders of scattering matrix including biexciton state. Firstly, their result shows that the biexciton boundstate introduces resonance scattering for exciton interaction with anti-parallel spins. Second, the inclusion of the higher-order scattering matrices results in a strong modification of the exciton-exciton interaction both for parallel and anti-parallel excitons (called "continuum correlations" \cite{Axt1998,Kwong2001a,Takayama2002,Oestreich1995a}). This continuum correlations might microscopically explain the necessity of the exciton-exciton interaction with anti-parallel spins $g_{\scriptscriptstyle +-}$, but the investigation of the exciton-exciton interaction starting from electron-hole basis is beyond the scope of our paper.\\  
In our model, the phase space filling effect is omitted due to its weak contribution. Based on the result of \cite{Ciuti2000},  the exchange interaction of excitons with parallel spins $g^{\rm theor}_{\scriptscriptstyle ++}$ and the contribution from phase space filling $g^{\rm theor}_{\rm PSF}$ can be estimated as $g^{\rm theor}_{\scriptscriptstyle ++}=3a_{\rm B}e^2/\epsilon_m S$ (in the Born approximation) and $g^{\rm theor}_{\rm PSF}=\Omega /n_{\mathrm sat}S$. Here $e$ and $\epsilon_m$ respectively represent elementary charge and dielectric constant of the quantum wells. $S$ is a quantization area. Considering a Bohr radius $a_{\rm B}$=12 nm, a dielectric constant $\epsilon_m=13.9\epsilon_0$ ($\epsilon_0$ is the vacuum permittivity) \cite{GrassiAlessi2000}, $\Omega $=3.26 meV and exciton saturation density $n_{\rm sat}\simeq 1\times 10^{11}$ cm$^{-2}$, we obtain a ratio $g^{\rm theor}_{\rm PSF}/g^{\rm theor}_{\scriptscriptstyle ++}\simeq 0.07$. This ratio indicates that the Coulomb term is the dominant contribution to the repulsive interaction of polaritons with parallel spins in our sample.\\
Using the Heisenberg equation of motion and mean field approximation \cite{Ciuti2000}, we obtain the equations of motion for the exciton, photon and biexciton. In this Hamiltonian, biexcitons are created from spin-up and spin-down excitons, which indicates that counter-circular polarization configuration is mandatory to induce the biexciton effect \cite{Wouters2007,Kuwata-Gonokami2000,Ivanov1993,Ivanov1995}. The exciton, photon and biexciton wavefunctions are described by the following exciton-photon Gross-Pitaevskii equation system:
%%\begin{equation}
%%i\hbar \dot{\psi}_{x,\uparrow}=(\epsilon_x-i\frac{\gamma_x}{2})\psi_{x,\uparrow}+\Omega\psi_{c,\uparrow}%%+g_{\tiny ++}|\psi_{x,\uparrow}|^2\psi_{x,\uparrow}+g_{\tiny +-}|\psi_{x,\downarrow}|^2\psi_{x,\uparrow}
%%\label{eq_cxgp1}
%%\end{equation}
\begin{subequations}
\begin{equation}\label{eq_cxgp1}
\begin{split}
i\hbar \dot{\psi}_{x,\uparrow} & =(\epsilon_x-i\frac{\gamma_x}{2})\psi_{x,\uparrow}+\Omega\psi_{c,\uparrow}\\
&\quad +2g_{\scriptscriptstyle ++}|\psi_{x,\uparrow}|^2\psi_{x,\uparrow}+g_{\scriptscriptstyle +-}|\psi_{x,\downarrow}|^2\psi_{x,\uparrow}\\ & \hspace{120pt} +g_{bx}\psi_B\psi_{x,\downarrow}^*
\end{split}
\end{equation}
\begin{equation}
i\hbar \dot{\psi}_{c,\uparrow}=(\epsilon_c-i\frac{\gamma_c}{2})\psi_{c,\uparrow}+\Omega\psi_{x,\uparrow}-F
\label{eq_cxgp2}
\end{equation}
\begin{equation}
i\hbar \dot{\psi}_{B}=(\epsilon_B-i\frac{\gamma_B}{2})\psi_{B}+g_{bx}\psi_{x,\uparrow}\psi_{x,\downarrow}.
\label{eq_cxgp3}
\end{equation}
\end{subequations}
$\gamma_x$, $\gamma_c$ and $\gamma_B$ are respectively exciton, photon and biexciton decay rate. In the numerical calculation, we use $\gamma_x=\gamma_c=0.53$ meV. $F$ represents an external source of photons given by the laser pulses.   

\subsection{A. Parallel spin polariton interaction}
In the co-circular polarization configuration, only the $g_{\scriptscriptstyle ++}$ interaction term is involved. We abbreviate the spin index since all polariton spins are the same. Our co-circular polarization pump-probe configuration is equivalent to that of $\chi^{(3)}$ parametric amplification. Thus, if we assume that the pump and probe beams have momentum $k_{pu}=0$ and $k_{pr}$ respectively, an idler beam appears with a momentum $k_{i}=2k_{pu}-k_{pr}=-k_{pr}$. Since in the experiment $k_{pr}\simeq 0$, we neglect the dispersion of photons. We substitute the wavefunction:
\begin{equation}
\psi_{x(c)}=\psi_{x(c)}^{pu}+\psi_{x(c)}^{pr}e^{ikx}+\psi_{x(c)}^{i}e^{-ikx}
\label{eq_psi}
\end{equation}
into the exciton-photon Gross-Pitaevskii equations (\ref{eq_cxgp1}) and (\ref{eq_cxgp2}). Assuming $|\psi_{x(c)}^{pu}|>>|\psi_{x(c)}^{pr,i}|$ the feedback on the pump wavefunction from the signal and idler can be discarded, thus the pump wavefunction $\psi_{x(c)}^{pu}$ can be written as equations (\ref{eq_cxgp1}) and (\ref{eq_cxgp2}). Additionally, we neglect the terms that do not conserve momentum. The dynamics of the set of wavefunctions $\vec{u}=(\psi^{pr}_{x},\psi^{pr}_{c},\psi^{i*}_{x},\psi^{i*}_{c})$ \cite{Lecomte2014} can be described as $i\hbar\dot{\vec{u}}=M^{\scriptscriptstyle ++}\vec{u}-\vec{F}^{pr}$. The matrix $M^{\scriptscriptstyle ++}$ is given by 
\begin{widetext} 
\begin{equation}
M^{\scriptscriptstyle ++}
=
\left( \begin{array}{cccc}
\epsilon_x + 2g_{\scriptscriptstyle ++}|\psi^{pu}_x|^2 -i\gamma_x/2 & \Omega & g_{\scriptscriptstyle ++}\psi_x^{pu 2} & 0 \\
\Omega & \epsilon_c-i\gamma_c/2 & 0 & 0 \\
g_{\scriptscriptstyle ++}\psi_x^{pu* 2} & 0 & \epsilon_x + 2g_{\scriptscriptstyle ++}|\psi^{pu}_x|^2 -i\gamma_x/2 & \Omega \\
0 & 0 & \Omega & \epsilon_c-i\gamma_c/2 \\
\end{array} \right)
\end{equation} 
\end{widetext}
Since the lower and upper polaritons are excited with spectrally broad femtosecond pulse, we model the pump and probe photon pulses as instantaneous delta function pulses exciting the system simultaneously. By Fourier transforming the temporal photon probe wavefunctions,  we obtain the spectra of the light emitted out of the cavity. Subtracting the pump-probe spectra from the reference spectra, we single out the energy shift of both lower and upper polariton resonances due to the presence of the pump pulse. The strength of our method to extract spinor interaction constants is based on the joint comparison of lower and upper polariton energy shifts.\\

\begin{figure*}
%\begin{center}
%\includegraphics[width=13cm,bb=0 0 2101 1109]{3.png}
\includegraphics[width=0.65\textwidth]{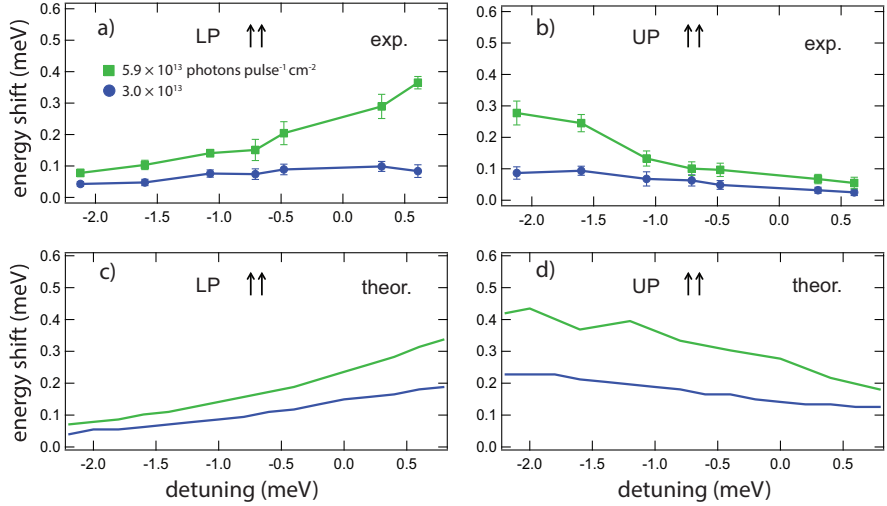}
\caption{(Color online) Experimental lower (a) and upper (b) polariton energy shift as a function of the cavity detuning for parallel spin polaritons. Simulation of lower (c) and upper (d) polariton energy shift based on exciton-photon basis Hamiltonian. The blue and green line respectively represents the experimental (simulated) pump photon density: $3.0\times 10^{13}$ ($|F^{pu}|^2=1$) and $5.9\times 10^{13}$ ($|F^{pu}|^2=2$) photons pulse$^{-1}$ cm$^{-2}$. $|F^{pu}|^2$ is a normalized excitation pump photon density}\label{expe_sim_co}

%\end{center}
\end{figure*}

In Figure \ref{expe_sim_co}(a) and (b), we plot respectively the experimental data of lower and upper polariton energy shifts as a function of the cavity detuning, for pump photon densities of $3.0\times 10^{13}$ and $5.9\times 10^{13}$ photons pulse$^{-1}$ cm$^{-2}$. We obtain the parallel spin interaction constant $g_{\scriptscriptstyle ++}=1$ meV$/n_0$ by simulating the lower and upper polariton energy shifts for the two pump powers (Figure 4(c) and (d)). Here, $n_0$ is a normalization particle density for the excitation pump photon density $|F^{pu}|^2=1$. The experimental results are well reproduced: The lower polaritons show an increase of the energy shift from negative to positive cavity detuning, while the upper polariton’s energy shift decreases. This behavior highlights the fact that lower polariton becomes more excitonic and upper polariton becomes more photonic for cavity detuning ranging from negative to positive values.\\

\subsection{B. Anti-parallel spin polariton interaction}
In the counter-circular polarization configuration, the large population of spin-up polaritons is probed with a small population of spin-down polaritons. Unlike the co-circular polarization configuration, the idler beam does not appear since the spin momentum conservation is  not satisfied \cite{Yaffe1993}.\\
The dynamics of the spin-down probe can be described using three coupled equations of motion: $i\hbar\dot{\vec{u}}=M^{\scriptscriptstyle +-}\vec{u}-\vec{F}^{pr}$, where the vector $\vec{u}=(\psi^{pr}_{x \downarrow},\psi^{pr}_{c \downarrow},\psi_B)$ is composed of the probe exciton, probe photon and biexciton wavefunctions respectively. The matrix $M^{\scriptscriptstyle +-}$ is given by
\begin{equation}
\begin{split}
&M^{\scriptscriptstyle +-}
=\\
&\left( \begin{array}{ccc}
\epsilon_x + g_{\scriptscriptstyle +-}|\psi^{pu}_{x\uparrow}|^2 -i\gamma_x/2 & \Omega & g_{bx}\psi^{pu *}_{x\uparrow} \\
\Omega & \epsilon_c-i\gamma_c/2 & 0 \\
g_{bx}\psi^{pu}_{x\uparrow} & 0 & \epsilon_{B}-i\gamma_B/2\\
\end{array} \right)
\end{split}
\end{equation}
   
\begin{figure*}
%\begin{center}
%\includegraphics[width=13cm,bb=0 0 2101 1109]{3.png}
\includegraphics[width=0.65\textwidth]{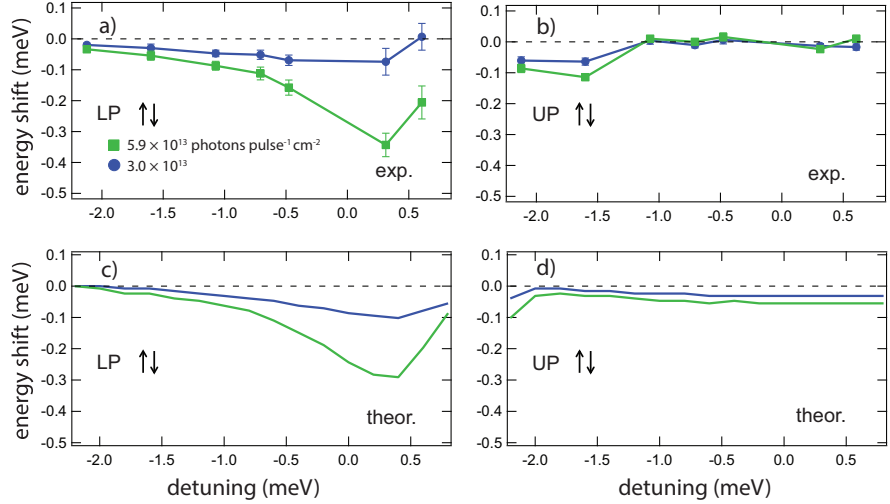}
\caption{(Color online) Experimental lower (a) and upper (b) polariton energy shift as a function of the cavity detuning for anti-parallel spin polaritons. Simulation of lower (c) and upper (d) polariton energy shift based on exciton-photon basis Hamiltonian. The blue and green line respectively represents the experimental (simulated) pump photon density: $3.0\times 10^{13}$ ($|F^{pu}|^2=1$) and $5.9\times 10^{13}$ ($|F^{pu}|^2=2$) photons pulse$^{-1}$ cm$^{-2}$. }\label{expe_sim_cross}
%\end{center}
\end{figure*}

In the matrix, $\psi^{pu}_{x(c)\uparrow}$ represents pump exciton (photon) wavefunction and follows the equation of motion (\ref{eq_cxgp1}) and (\ref{eq_cxgp2}). In the simulation, we set $E_B =2E_x-2.5$ meV and $\gamma_B=1.1$ meV. Similarly to the co-circular simulation, we obtain the interaction constant $g_{bx}=1.2$ meV$/\sqrt{n_0}$ and $g_{\scriptscriptstyle +-} = -1.2$ meV$/n_0$ from the comparison between simulation and experiment of the lower and upper polariton energy shifts at $3.0\times 10^{13}$ and $5.9\times 10^{13}$ photons pulse$^{-1}$ cm$^{-2}$ pump photon densities. We plot in Figure 5 the experimental and simulated lower and upper polariton energy shifts as a function of the cavity detuning. Similar to the experimental results, the simulation shows the enhancement of the lower polariton red shift energy from negative to positive cavity detuning (see Figure \ref{expe_sim_cross} (c)). This enhancement of the redshift of the lower polariton resonance comes from the scattering of lower polaritons (LP) via the biexciton state (BX) (LP-LP $\rightarrow$ BX). This effect plays an important role concerning the anti-parallel spin polariton interactions and should be considered in the analysis of the renormalization of the lower polariton energy. Notice that the polaritonic Feshbach biexciton resonance \cite{Takemura2014} is analogous to the Feshbach resonance in cold atoms \cite{Inouye1998,Timmermans1999} where the atom-atom interaction is altered due to a molecular bound state.\\
Concerning the upper polariton mode, the very small energy shift experimentally measured is well reproduced by the numerical simulation (Fig. \ref{expe_sim_cross}(d)). This behavior is specific of the counter-polarized configuration and then might be related to the presence of biexcitons. However, further investigations should be performed to deeply understand these observations.\\
In summary, our approach is more general than previously reported studies since we consider a simultaneous excitation of lower and upper polariton branches. We probe the energy shifts of both branches and, through comparison between the numerical and experimental results, we estimate the interaction constants of the exciton-photon-biexciton Hamiltonian (2) as $g_{\scriptscriptstyle ++}:g_{\scriptscriptstyle +-}:g_{bx}\simeq 1/n_0:-1.2/n_0:1.2/\sqrt{n_0}$ (meV). This ratio shows that the spinor polariton interaction with anti-parallel spins, which is usually considered to be very weak, is comparable to that of parallel spins. However it is very important to note that, due to commutation relation, a factor of 2 appears in front of $g_{\scriptscriptstyle ++}$ (Eq. (\ref{eq_cxgp1})). As a consequence, the energy shift for parallel spins is twice the energy shift for anti-parallel one for the same density and cavity detuning. Additionally, the strong lower polariton energy shift with cavity detuning dependency appears due to the biexciton-exciton coupling. The joint study of lower and upper spinor polariton interaction described in this report, is well reproduced in the exciton-photon basis.\\
Usually, in resonant excitation experiments, only the lower polariton branch is excited. Then, instead of the excitonic interaction constants $g_{\scriptscriptstyle ++}$,  $g_{\scriptscriptstyle +-}$ and $g_{bx}$, the parallel $\alpha_1$ and anti-parallel $\alpha_2$ polariton-polariton interaction constants are defined within the lower-polariton basis. In order to validate and compare our results to other studies on polariton interactions, we describe below our observation in the lower polariton basis by changing the working basis from exciton-photon-biexciton to lower polariton basis.

\section{V Lower-polariton interaction constants}
When experiments with lower polaritons is considered, a continuous wave laser excitation is usually involved. In order to apply our extracted parameters to this configuration, we need to transform the exciton-photon basis Hamiltonian into lower-polariton Hamiltonian.\\
Firstly, we rewrite the Hamiltonian (\ref{eq_H}) using lower-polariton basis defined as $\hat{p}_{\rm lp}=X\hat{x}+C\hat{c}$ and neglect terms that includes upper-polaritons \cite{Ciuti2000}:
\begin{equation}\label{Hlp}
\begin{split}
\hat{H}&=\epsilon_c\hat{c}^{\dagger}_{\uparrow}\hat{c}_{\uparrow}+\epsilon_x\hat{x}^{\dagger}_{\uparrow}\hat{x}_{\uparrow}+\Omega(\hat{c}^{\dagger}_{\uparrow}\hat{x}_{\uparrow}+\hat{x}^{\dagger}_{\uparrow}\hat{c}_{\uparrow})+\hat{H}_{\rm int}\\ 
&\simeq \epsilon_{lp}\hat{p}_{\rm lp}^{\dagger}\hat{p}_{\rm lp}\\ 
&+g_{\scriptscriptstyle ++}|X|^4\hat{p}_{lp, \uparrow}^{\dagger}\hat{p}_{lp, \uparrow}^{\dagger}\hat{p}_{lp, \uparrow}\hat{p}_{lp, \uparrow}\\ 
&+g_{\scriptscriptstyle +-}|X|^4\hat{p}_{lp, \uparrow}^{\dagger}\hat{p}_{lp, \downarrow}^{\dagger}\hat{p}_{lp, \downarrow}\hat{p}_{lp, \uparrow}\\ 
&+g_{bx}(X^2\hat{p}_{lp,\uparrow}\hat{p}_{lp,\downarrow}\hat{B}^{\dagger}+X^{* 2}\hat{p}_{lp,\uparrow}^{\dagger}\hat{p}_{lp,\downarrow}^{\dagger}\hat{B})
\end{split}
\end{equation}
Here $X$ and $C$ are Hopfield coefficients defined as 
\begin{equation}
|X|^2=\frac{1}{2} \left(1+\frac{\delta}{\sqrt{\delta^2+\Omega^2}}\right)\label{Hopfield1}
\end{equation}
and
\begin{equation}
\displaystyle
|C|^2=\frac{1}{2} \left(1-\frac{\delta}{\sqrt{\delta^2+\Omega^2}}\right).\label{Hopfield2}
\end{equation}

The lower polariton energy $\epsilon_{lp}$ is calculated as 

\begin{equation}
\epsilon_{lp} = \frac{\epsilon_c+\epsilon_x}{2}-\frac{1}{2}\sqrt{(\epsilon_c-\epsilon_x)^2+\Omega^2}.
\end{equation}

Based on the Hamiltonian (\ref{Hlp}), we calculate the Heisenberg equations of motion applying mean-field approximation:
\begin{equation}
\begin{split}
i\hbar \dot{\psi}_{lp, \uparrow}&=(\epsilon_{lp}-i\frac{\gamma_{lp}}{2})\psi_{lp, \uparrow}+2g_{\scriptscriptstyle ++}|X|^4|\psi_{lp, \uparrow}|^2\psi_{lp, \uparrow}\\
& +g_{\scriptscriptstyle +-}|X|^4|\psi_{lp, \downarrow}|^2\psi_{lp, \uparrow}+g_{bx}X^{*2}\psi^*_{lp, \downarrow}\psi_{B}-F
\end{split}\label{10}
\end{equation}
\begin{equation}\label{11}
i\hbar \dot{\psi}_{B}=(\epsilon_B-i\frac{\gamma_{B}}{2})\psi_{B}+g_{bx}X^2\psi_{lp, \uparrow}\psi_{lp, \downarrow}
\end{equation}
where $\gamma_{lp}=|C|^2\gamma_c + |{\rm X}|^2\gamma_{\rm x} $ is the lower polariton decay rate.\\
In the following, we address equations (\ref{10}) and (\ref{11}) in parallel (A) and anti-parallel (B) spin polariton frameworks.

\subsection{A. Parallel spin polariton interaction}
For the parallel spin configuration, the dynamics of the $\psi_{lp}$ wave function is given by 
\begin{equation}
i\hbar \dot{\psi}_{lp}=(\epsilon_{lp}-i\frac{\gamma_{lp}}{2})\psi_{lp}+2g_{\scriptscriptstyle ++}|X|^4|\psi_{lp}|^2\psi_{lp}-F
\end{equation}
Thus, the mean-field energy shift can be expressed as
\begin{equation}\label{DeltaEplus}
\Delta E_{++}=\alpha_1 |\psi_{lp}|^2 
\end{equation} 
where $|\psi_{lp}|^2$ is the lower polariton density. The lower polariton interaction with parallel spins is thus defined as
\begin{equation}\label{alphaone}
\alpha_1=2g_{\scriptscriptstyle ++}|X|^4.
\end{equation}
On Figure \ref{stationary} (a), we plot the energy shift of the lower polariton for parallel spin interaction (eq. (\ref{DeltaEplus})) using the value of $g_{\scriptscriptstyle ++}$ extracted previously and pump polariton density $|\psi_{lp}|^2=0.2$. The energy shift follows a dependence given by the excitonic content of the lower polariton through the term $|X|^4$ as reported before by Ciuti \emph{et al.}\citep{Ciuti2000}. 

\subsection{B. Anti-parallel spin polariton interaction}

\begin{figure}
\begin{center}
\includegraphics[width=0.4\textwidth]{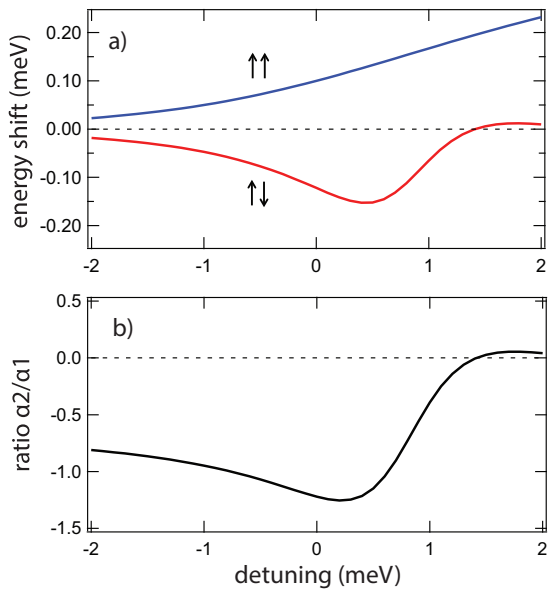} 
\caption{(Color online) (a) The energy shift $\Delta E_{++}$ (blue) and $\Delta E_{+-}$ (red) obtained in the equation (\ref{DeltaEplus}) and (\ref{DeltaEminus}) as a function of the cavity detuning for the pump polariton density $|\psi_{lp, \uparrow}|^2=0.2$. (b) The ratio $\alpha_2/\alpha_1$ as a function of the cavity detuning}\label{stationary}
\end{center}
\end{figure} 
In order to obtain the lower polariton interaction with anti-parallel spins, we consider pump-probe spectroscopy in the counter-circular polarization configuration. Assuming that the spin-up pump pulse is much stronger than the spin-down probe pulse, we can express the probe lower polariton dynamics as $i\hbar\dot{\vec{u}}=M\vec{u}-\vec{F}^{pr}$, where the vector $\vec{u}$ is $\vec{u}=(\psi^{pr}_{lp\downarrow},\psi_{B})$ and $\vec{F}^{pr}$ is a source term. The matrix $M$ is 
\begin{equation}
\begin{split}\displaystyle
M
&=\\
&\left( \displaystyle\begin{array}{cc}
\epsilon_{lp} + g_{\scriptscriptstyle +-}|X|^4|\psi^{pu}_{lp\uparrow}|^2 -i\gamma_{lp}/2 & ~~~~ g_{bx}X^{*2}\psi^{pu *}_{lp\uparrow} \\[0.8em]
g_{bx}X^{2}\psi^{pu}_{lp\uparrow} & \epsilon_{B}-i\gamma_B/2\\
\end{array} \right)
\end{split}
\end{equation}
Let us consider that the pump is resonant with the lower-polariton energy $\psi^{pu}_{lp\uparrow}=|\psi^{pu}_{lp\uparrow}|e^{-i \epsilon_{lp}t/\hbar}$ and probe with energy $\epsilon$: $\vec{F}^{pr}=(1,0)e^{-i\epsilon t/\hbar}$. For the steady state solutions, we assume that the pump wavefunction has a form $\psi^{pu}_{lp\downarrow}=\psi^{pu}_{lp\downarrow}(\epsilon)e^{-i\epsilon_{lp} t/\hbar}$ and the biexciton wavefunction is expressed as $\psi_{B}=\psi_{B}(\epsilon)e^{-i(\epsilon+\epsilon_{lp})t/\hbar}$. Replacing $\psi^{pu}_{lp\uparrow}$ and $\psi^{pu *}_{lp\uparrow}$ with $|\psi^{pu}_{lp\uparrow}|$ in the matrix $M$, we can obtain the steady-state solution of the probe lower polariton spectrum as
\begin{equation}
\psi^{pr}_{lp\downarrow}(\epsilon)
=(1 \ 0)
\left[ M-\left( \begin{array}{c c}
\epsilon & 0\\
0 & \epsilon_{lp}+\epsilon\\
\end{array} \right)\right]^{-1}
\left( \begin{array}{c}
1 \\
0 \\
\end{array} \right)
\end{equation}
The analytic solution of $\psi^{pr}_{lp\downarrow}(\epsilon)$ is 
\begin{eqnarray}\displaystyle\label{prwavefunction}
\psi^{pr}_{lp\downarrow}(\epsilon)=\biggl[ \epsilon_{lp} -\epsilon\ + \ &g_{\scriptscriptstyle +-}&|X|^4|\psi^{pu}_{lp\uparrow}|^2 -i\gamma_{lp}/2 \nonumber\\
 &-& \frac{g_{bx}^2|X|^4|\psi^{pu}_{lp\uparrow}|^2}{\epsilon_{B}-\epsilon_{lp}-\epsilon-i\gamma_B/2} \biggr]^{-1}\nonumber\\
\end{eqnarray}
Under the assumption of a weak exciton-biexciton coupling and a low pump excitation density ($g_{bx}|\psi^{pu}_{lp\uparrow}|<\gamma_B/2$), the two-mode solutions of $\psi^{pr}_{lp\downarrow}(\epsilon)$  might be approximated by a single solution. Indeed, substituting $\epsilon=\epsilon_{lp}$ and taking the real-part of the eq. (\ref{prwavefunction}), the pump induced energy shift $\Delta E_{+-}$ can be approximated as 

\begin{equation}\label{DeltaEminus}
\Delta E_{+-}\simeq g_{\scriptscriptstyle +-}|X|^4|\psi^{pu}_{lp\uparrow}|^2
 -\frac{g_{bx}^2|X|^4|\psi^{pu}_{lp\uparrow}|^2(\epsilon_{B}-2\epsilon_{lp})}{(\epsilon_{B}-2\epsilon_{lp})^2+(\gamma_B/2)^2} 
\end{equation}
and the lower polariton interaction with anti-parallel spins is defined as 

\begin{equation}\label{alphatwo}
\alpha_2\simeq g_{\scriptscriptstyle +-}|X|^4
 -\frac{g_{bx}^2|X|^4(\epsilon_{B}-2\epsilon_{lp})}{(\epsilon_{B}-2\epsilon_{lp})^2+(\gamma_B/2)^2}. 
\end{equation}
This result shows the two contributions to the energy shift in the counter-polarized configuration that bring strong similarity to derived expressions for optical Feshbach resonance in cold atoms \cite{Theis2004}. First, the term proportional to the excitonic content $|X|^4$ related to anti-parallel spin polariton interaction is usually called background interaction term. Second, the lower-polariton-biexciton coupling term takes the form of a Lorentzian profile. Figure \ref{stationary}(b) displays the behavior of the energy shift in the counter-polarized configuration considering the value of $g_{\scriptscriptstyle +-}$ and $g_{bx}$ extracted from the numerical simulations and pump polariton density $|\psi_{lp, \uparrow}|^2=0.2$. The energy shift displays a change in sign and amplitude when the energy of two-lower polariton crosses the energy of the biexciton, which here occurs at $\delta\approx1$ meV \cite{Takemura2014}.\\
On Figure \ref{stationary}(b), we plot the ratio between counter and co-polarized interactions. We obtain a ratio as the one reported previously using a different approach \cite{Vladimirova2010}. For negative detuning the ratio takes a negative value, -0.8, that progressively increases until reaching -1.2, then dropping suddenly to a sligthly positive value. This comparison to previous results strengthens the interest and generality of the method presented here.
\section{VI. CONCLUSION}
In conclusion, we provided a direct demonstration that the lower and upper polaritons with parallel spins interact repulsively and attractively when possessing anti-parallel spins. In addition, by modeling the experiments with the exciton-photon based Gross-Pitaevskii equation we extracted the spinor microscopic exciton interaction strengths. The joint analysis of the lower and upper polariton energy renormalization with cavity detuning reveals the dominant role of the biexciton in anti-parallel spin configuration. Through the lower polariton basis model, we validate our results and in addition, we give the analytic expression for the polariton interaction with parallel spin $\alpha_1$ and with anti-parallel spin $\alpha_2$ with the explicit role the biexcitonic resonance effect.\\
\newline
The present work is supported by the Swiss National Science Foundation under project N◦135003, the Quantum Photonics National Center of Competence in research N115509 and from European Research Council under project Polaritonics contract N◦219120. The POLATOM Network is also acknowledged.

%\bibliography{biblio}

%\bibliography{bibfileV35}
%\begin{thebibliography}{9}

%\end{thebibliography}

\end{document}